\documentclass[11pt,a4paper]{article}

\usepackage[utf8]{inputenc}
\usepackage[T1]{fontenc}
\usepackage{lmodern}
\usepackage{microtype}
\usepackage[margin=1.1in]{geometry}
\usepackage{setspace}
\usepackage{amsmath,amssymb}
\usepackage{natbib}
\usepackage{booktabs}
\usepackage{hyperref}

\hypersetup{
  colorlinks=true,
  linkcolor=black,
  citecolor=black,
  urlcolor=black
}

\title{The frame problem in quantitative practice:\\[0.3em]
ontological uncertainty and epistemic humility in an age of automated inference}

\author{William Fauriat\\[0.4em]
\small CEA, DAM, DIF, F-91297, Arpajon, France\\[0.2em]
\small \texttt{william.fauriat@cea.fr}}

\date{}

\begin{document}
\maketitle
\begin{abstract}
\noindent Quantitative practice across statistics, engineering, and machine
learning has been transformed by the automation of inference. Predictions are
produced, validated, and deployed at scale and speed that human-mediated
reasoning could not match. This shift intersects with a structural limit of
reasoning that no methodological refinement dissolves: every inference rests
on a \emph{finite specification} of conditions, and what falls outside the
specification does not appear as a widened uncertainty band --- it does not
appear at all. The choice of specification --- the \emph{frame} --- is
upstream of the inference and cannot be audited from inside the system that
uses it. This paper offers a synthetic, application-oriented review. We
argue that three categories of uncertainty operate in quantitative practice
--- aleatory, epistemic, and \emph{frame} (or \emph{ontological}) --- and that the third, the
residue of finite specification, is structurally invisible to formal
analysis within the chosen frame and is the locus of most consequential
failures. We trace why the limit applies equally to deductive and inductive
reasoning, why no meta-level procedure dissolves the regress, and why
current conditions of automated inference make \emph{epistemic humility}
--- the practical disposition this argument supports --- more, not less,
important. We articulate the argument's specific resonances for five
typical figures of contemporary quantitative work --- the engineer, the
statistician, the mathematician, the machine-learning practitioner, and
the non-specialist recipient of expert claims --- showing how the
structural argument bears on each practice's natural defenses. The argument
is not against rigor or against quantification; it is for distinguishing
rigor earned within a frame from rigor with respect to the frame.

\vspace{0.6em}
\noindent\textbf{Keywords:} epistemic humility; frame problem; ontological
uncertainty; automated inference; uncertainty quantification; machine
learning; statistical decision theory; conformal prediction; model adequacy.
\end{abstract}

\vspace{0.8em}
\noindent\textbf{Note on production.} The author has chosen to inform
readers explicitly of the following. (1)~The prose of this article has
been generated by a large language model (Anthropic's Claude, Opus
family). (2)~The theoretical substrate of the argument is largely the
product of the referenced literature, which the author acknowledges,
and is continuous with a perspective the author developed earlier
in~\citet{Fauriat2022} --- i.e., predating any model-mediated production.
Together with its practical implications,
that substrate has been curated across a large number of iterations
with the model, including multiple intermediate productions of various lengths
and technical depth --- short essays, long presentations, and cross-critique literature reviews.
The current article is the result of this slow
and meticulous filtering, which the author considers to constitute
acceptable research work. (3)~The wide, cross-disciplinary reach of
the synthesis would not, in the acknowledgment of the author, have
been possible without the instrumental support of the language
model, deployed throughout as a literature explorer and an aid to structuring complex
reasoning chains,
always in directions chosen by the author, according to the author's own perspective,
which the article illustrates.
(4)~The author owns and stands by the final --- curated and revised
--- version, and invites the reader to judge the result by its
content; reservations about the production process are understandable
and are acknowledged by the author. (5)~The author hopes the importance of the message --- from the author's
perspective --- will weigh in the reader's judgment alongside any
reservations about how it was produced. (6)~The author is aware of the
irony implied by the following stance: warning against excessive automation of
reasoning, while using tools at the frontier of automation. The author would point
out that this paradox is not in contradiction with the author's position:
understand limitations in tool use with a self-critique disposition, neither dismiss
automation nor let it roam unchecked.

\vspace{1.5em}

\section{Introduction}
\label{sec:intro}

The Challenger O-rings did not fail because the engineering analysis was
imprecise. They failed because the model class within which the analysis
was conducted did not include, with sufficient salience, the dependence of
O-ring resilience on launch-day temperature outside the calibrated range
\citep{Vaughan1996}. The pre-2008 financial risk models did not produce a
wide confidence interval on joint failure across geographically distributed
mortgage pools; they did not produce that scenario at all
\citep{KingKay2020}. A modern deep network fine-tuned to high accuracy on a
benchmark distribution does not flag, with calibrated uncertainty, inputs
drawn from a distribution it has never seen; on those inputs, its
calibration apparatus has nothing to say
\citep{QuioneroCandela2009,Amodei2016}. In each case, what failed was not
internal rigor. What failed was the frame.

This paper is a synthetic review of a structural feature of quantitative
inference common to such failures. The feature has been named in several
disciplinary vocabularies. \citet{Knight1921} distinguished measurable
risk from unmeasurable \emph{uncertainty}, identifying a class of
situations in which the probability space itself is not given.
\citet{Box1976,Box1979} compressed half a century of statistical practice
into the maxim that all models are wrong but some are useful, with an
implication often underweighted: every model is wrong in a \emph{specific}
way determined by the frame chosen. The artificial-intelligence literature
introduced the \emph{frame problem} as a structural feature of any
reasoning system that must filter relevance from a finite specification
\citep{McCarthyHayes1969,Dennett1984,Pylyshyn1987,Shanahan2016}. The
statistical literature on robustness, sensitivity, and prior elicitation
has long acknowledged the gap between within-model analysis and adequacy
of the model class
\citep{Berger1985,Saltelli2008,DraperJRSS1995,HoetingBMA1999}. The
engineering literature on verification, validation, and uncertainty
quantification has, over the last quarter-century, formalized the partition
between aleatory and epistemic uncertainty
\citep{HeltonOberkampf2004,OberkampfRoy2010,Walker2003}. The
machine-learning literature has named distribution shift, dataset shift,
specification gaming, and reward hacking as operational symptoms of the
same structural feature
\citep{QuioneroCandela2009,Amodei2016,Christian2020,HullermeierWaegeman2021}.

Each of these literatures has identified the problem in its own
vocabulary. They have not always recognized that they are addressing the
same structural object. The first aim of this paper is to lay that object
out plainly. The second is to address its specific bearing on the
heterogeneous practices that compose contemporary quantitative work ---
engineers integrating heterogeneous models in safety-critical pipelines,
statisticians designing workflows whose conditions are silently relaxed at
deployment, mathematicians whose formal tools are routinely deployed under
metaphysical commitments the formalism does not entail,
machine-learning practitioners whose intuition runs against any suggestion
that scale fails to dissolve old problems, and non-specialist
decision-makers and citizens who are increasingly the consumers of
confident outputs whose conditions they cannot inspect. A review that does
not address this heterogeneity will be correctly received by each audience
as not about them.

The structural argument has the shape of a logical arc with an operational
coda. The arc, developed in Sections~\ref{sec:picture}--\ref{sec:nfl},
establishes that every prediction works on a finite specification of
conditions chosen upstream of the inference, that this limit applies
equally to deductive and inductive modes, that a reasoning system cannot
audit its own frame from within, and that no meta-level procedure
dissolves the regress without itself becoming a frame.
Section~\ref{sec:automation} provides the coda: current conditions of
automated inference make the structural limit operationally urgent rather
than merely structurally true. Section~\ref{sec:practices} then walks the
argument through the five practices named above, identifying for each
the legitimate entry point, the natural defense, and the structural
counter-move. Section~\ref{sec:humility} articulates the disposition the
argument earns --- \emph{epistemic humility} as a practical posture
rather than a stylistic preference --- and Section~\ref{sec:discussion}
develops its implications for builders, decision-makers, institutions,
and pedagogy.

\section{Acting through a constructed picture}
\label{sec:picture}

The basic structure of any rational endeavor is a target, a world, and the
gap between them \citep{Simon1982}. To close the gap is to redirect a
fragment of the world toward a preferred state, which requires
understanding how the relevant parts of the world work. That understanding
is never direct. It is always mediated --- by perception, by measurement,
by models, by inference. We do not touch the world; we touch our
representation of it \citep{Korzybski1933}. Every decision, however
data-rich, ultimately rests on a constructed picture of what matters and
how the matters connect. That construction is what the agent acts on. The
world, however, responds to the action, not to the picture.

The asymmetry between picture and action is not a sceptical claim about
whether the world can be known; it is a structural observation about where
the agent stands. The question this paper develops is what happens when
actions land in a world that does not consult the picture that informed
them, and what disposition that asymmetry warrants for the agent who must
act anyway.

A short illustration. An agent considers crossing a sea in a small boat
and weighs whether to take a satellite phone. To reason about the choice,
the agent assembles a picture: weather, equipment, season, skill, the
distribution of likely incidents. The agent attaches a probability ---
twenty percent, say, of needing the phone. The crossing then happens. It
is not, in the event, a twenty-percent crossing. The phone is used, or it
is not. The world resolves to a single outcome along an axis the model
assigned a distribution to. The probability did not describe the world;
it described the agent's judgment about the world, given a frame the
agent had chosen. The crossing responded to the action --- bringing the
phone or not --- not to the probability.

The machinery of probability, prediction, and inference is therefore best
understood as machinery of judgment under a chosen frame, not as machinery
of direct access to the world. We make this point precise in
Sections~\ref{sec:frame} and~\ref{sec:formality}. The philosophical
ancestor of the framing is Hume's problem of induction \citep{Hume1748};
the modern statistical articulations descend, among others, from
\citet{deFinetti1974}, \citet{Savage1954}, \citet{Jaynes2003},
\citet{Lindley2006}, and \citet{Pearl1988}.

\section{The frame problem in quantitative reasoning}
\label{sec:frame}

\subsection{Relevance and conditioning}

Any reasoning system, human or automated, must first decide what is
relevant. Which variables to track, which interactions to model, which
data to collect, which conditioning to perform. The decision is upstream
of any inference. Relevance, however, cannot be determined from nowhere.
It is always determined from somewhere: from prior knowledge, prior
experience, prior model class. Perception shapes the frame; the frame
shapes perception \citep{McCarthyHayes1969,Dennett1984}. The loop is
closed and cannot be opened from inside the system that uses it.

The statistical formulation is direct. To predict a quantity $Y$ given a
context, the analyst specifies a conditioning set $X_1, X_2, \ldots, X_k$
and computes some function of $\Pr(Y \mid X_1, \ldots, X_k)$. The law of
total probability gives the structural form of prediction under
uncertainty,
\begin{equation}
\Pr(Y) \;=\; \sum_{x} \Pr(Y \mid X = x)\, \Pr(X = x),
\label{eq:total-prob}
\end{equation}
which can be read in two registers. The conditional distributions
$\Pr(Y \mid X = x)$ encode the modeling work --- the structure of how
contextually-conditioned outcomes are distributed. The marginal weights
$\Pr(X = x)$ encode the assumptions --- how strongly each piece of context
is supported. Both are choices made within a frame that fixes which $X$s
are considered.

The frame problem is the question of how the analyst chooses
$X_1, \ldots, X_k$. The choice is circular: to decide which variables
matter, the analyst needs some idea of which variables could matter, which
presumes a prior characterization of the phenomenon, which is what the
analysis was supposed to help produce \citep{Pylyshyn1987,Shanahan2016}.
The circularity is not a defect of methodology but a structural feature
of reasoning under incomplete information. New relevance can enter the
system only from outside it --- by another vantage, another witness, or
through exploration \citep{SuttonBarto2018}; we return to this in
Section~\ref{sec:discussion}.

\subsection{Finite specification}

The conditioning set in any real analysis is finite. There are only so
many variables one can track, only so many interactions one can model,
only so much data one can collect. Finite specification engenders
uncertainty.

The structural consequence is the load-bearing claim of the paper:
\emph{what falls outside the specification does not appear as wider error
bars; it does not appear at all}. A confidence interval is computed within
a model; the model conditions on what was included; what was not included
is not in the interval. Robustness analysis that perturbs the included
variables remains within the included variables. Sensitivity analysis on
the prior moves probability within a distribution class; it does not relax
the class \citep{Berger1985,Saltelli2008}. Bayesian Model Averaging
\citep{HoetingBMA1999} extends inference over a specified class of
models, but the class is itself a choice. Conformal prediction guarantees
coverage of the calibration distribution; it does not guarantee that the
calibration distribution matches deployment
\citep{Vovk2005,AngelopoulosBates2023}.

The Russell--Hume turkey makes the point concrete
\citep{Russell1912,Hume1748,Taleb2007}. A turkey, observing that it is fed
each morning, fits an inductive model. The model has impeccable internal
calibration: its predictive confidence grows monotonically with each
confirming observation, exactly as a sound inductive procedure
prescribes. On the morning before the holiday, the model's confidence is
at its peak. The failure that follows is not in the turkey's statistics.
It is in the frame within which the statistics were computed. Nothing in
the within-frame data could have signalled that the frame was inadequate,
because the within-frame data was generated under the frame. The surprise
came from outside it --- from a structure the model had no category for.

This is the structural reason a model can pass every within-frame
quantitative check and still fail. The checks operate within the frame;
the failure comes from outside it.

\section{Three kinds of uncertainty}
\label{sec:three}

The recognition that finite specification is uncertainty's source
recommends a taxonomy on which the literature has gradually converged,
though not under a uniform label.

The familiar pair separates \emph{aleatory} uncertainty --- variability
the analyst represents as irreducible within the present analysis,
typically by distribution --- from \emph{epistemic} uncertainty --- the
component that more data, better measurement, or finer modeling could in
principle shrink. The labels are functional rather than metaphysical:
what is treated as aleatory at one level of analysis can be re-specified
as epistemic at a finer one, and \emph{it is the boundary between the two
that the analyst chooses}, not the irreducibility of either. The
distinction is a modeling decision about where to stop refining, not a
property of the world. The pair is well established in engineering
uncertainty quantification
\citep{HeltonOberkampf2004,OberkampfRoy2010} and is widely used in risk
analysis \citep{MorganHenrion1990}. Aleatory components are typically
modelled with distributions and propagated forward; epistemic components
are reduced by acquiring additional information or by improving the
model class. Recent machine-learning work has adopted the same partition
for predictive uncertainty in deep models
\citep{KendallGal2017,HullermeierWaegeman2021}.

A third category sits upstream of these two. The practical literature has
named it variously: \emph{ontological} uncertainty
\citep{Wynne1992,Walker2003,FuntowiczRavetz1990}, model-form or
model-structure uncertainty \citep{DraperJRSS1995,HoetingBMA1999},
``deep'' uncertainty in policy analysis \citep{Lempert2003}, and
\emph{Knightian} uncertainty in the economics tradition
\citep{Knight1921}. The differences in usage matter for specialists; the
shared object is the same: uncertainty about whether the frame itself is
adequate. Not ``I do not know the value of $X$,'' but ``I am not sure
$X$, as I have defined it, is the right object to look at.'' Not ``the
model has wide error bars on this prediction,'' but ``the model's
categories may not match the structure of the problem.''

This third category is the residue of finite specification, and it is
what within-frame methodology cannot address from inside the frame.
Sensitivity analysis perturbs within a model class; robustness analysis
bounds the residue inside an assumed class; calibration is computed
against a reference class that is itself a choice; Bayesian Model
Averaging averages over a specified collection of models. Each of these
tools is sound and valuable within its scope. None addresses the adequacy
of the frame, because each operates inside one.

The asymmetry matters. The first two categories are \emph{within-frame}
uncertainties; the third is the residue of the frame choice itself.
Within-frame uncertainties admit formal treatment. The third does not, in
the strict sense, admit formal treatment --- not because the discipline
has failed to invent the right formalism, but because any formal closure
of the third category would itself be a frame. We return to this point as
the meta-level no-free-lunch in Section~\ref{sec:nfl}.
Table~\ref{tab:three-uncertainties} summarises the operative distinctions
and their typical methodological treatments.

\begin{table}[t]
\centering
\small
\begin{tabular}{p{0.16\textwidth} p{0.38\textwidth} p{0.38\textwidth}}
\toprule
\textbf{Category} & \textbf{Meaning} & \textbf{Typical treatment} \\
\midrule
Aleatory & Variability the analyst represents as irreducible within the
present analysis. The aleatory/epistemic boundary is itself a modeling
choice. & Probability distributions; Monte Carlo; ensemble generation. \\
\addlinespace
Epistemic & Reducible-with-information uncertainty about quantities
inside the frame. & Bayesian updating; data acquisition; model
refinement; cross-validation; predictive UQ in deep models. \\
\addlinespace
Ontological & Uncertainty about whether the frame --- variables,
categories, model class --- is adequate. & Not formally addressable from
inside the frame. Mitigations are dispositional: pre-mortem analysis,
adversarial review, scenario expansion, exploration, structural humility. \\
\bottomrule
\end{tabular}
\caption{Three categories of uncertainty in quantitative practice. The
first two admit formal treatment within the chosen frame. The third is
the residue of finite specification and is structurally invisible to
within-frame analysis.}
\label{tab:three-uncertainties}
\end{table}

\section{Formality does not lift the limit}
\label{sec:formality}

A natural move at this point is to suppose that the structural limit is a
feature of \emph{inductive} reasoning --- of pattern-fitting under finite
data --- and that \emph{deductive} reasoning, working from axioms by
necessity, escapes it. The move is intuitive, and it is wrong.

Two reasoning modes are worth distinguishing. \emph{Deduction} proceeds
from a description of conditions and rules to conclusions that follow with
necessity, given the description and the rules. The guarantee is
conditional: it holds if and only if the supporting description and rules
hold. \emph{Induction} proceeds from observations in some context to
generalisations about contexts that are sufficiently similar. The
guarantee is also conditional, but in a different way: it holds if the
deployment context is sufficiently similar to the observation context
and the observations are sufficiently solid.

Both guarantees are real. Both are conditional on a frame --- the
description in the deductive case, the similarity-of-context in the
inductive case. The frame problem applies to both: in deduction, the
choice of axioms, boundary conditions, and idealisations \emph{is} the
frame; in induction, the choice of features, conditioning variables, and
reference class \emph{is} the frame. Necessity inside a poorly-framed
problem is necessity-inside-a-poor-frame. The mathematics is sound; the
frame chose what entered the mathematics.

The contemporary causal-inference framework makes this particularly
explicit. The structural-equation or DAG-based account of causality
\citep{Pearl1988,Pearl2009,Hernan2020} requires that the analyst specify
which variables are nodes, which directional relations between them are
admitted, and which interventions are contemplated. The mathematical
machinery --- $d$-separation, the do-operator, identification theorems
--- then yields rigorous conclusions \emph{relative to the specified
causal frame}. The discipline is internally consistent and has produced
results that within-frame statistical methods alone cannot. But the
causal frame itself is not certified by the formalism; it is the prior
choice the formalism operates over. The framework has clarified, rather
than dissolved, the upstream choice.

Within the inductive mode, the same logic applies to modern formal
guarantees. Distribution-based predictive intervals are valid if and only
if the distributional hypothesis holds. Distribution-free methods such as
conformal prediction \citep{Vovk2005,AngelopoulosBates2023} provide valid
coverage if the deployment context is exchangeable with --- or otherwise
sufficiently similar to --- the calibration context. Either way, the
guarantee rests on a framing-level condition the method itself cannot
certify. The method is correct relative to its frame; the frame is not
certified by the method.

The deeper structural claim is that a reasoning system reasons from
within a frame and is structurally blind to its own blind spots. The
system can report calibrated confidence over the things it was trained or
constructed to consider. It has no representation of the things it was
not. Auditing the frame requires resources --- categories, observations,
salience --- that the frame does not carry. This is not a deficiency of
any particular system; it is a structural property of any reasoning
system that filters relevance from a finite specification.

A concrete statistical example. A clinical prediction model is trained
on a cohort with a particular covariate distribution. The model performs
well in internal validation and in external validation on cohorts
sufficiently similar to the training distribution. Deployed in a
population whose disease severity distribution has shifted --- whether
because the catchment area has changed, or because diagnostic practice
has changed --- the model degrades. The model was correct relative to its
frame; the frame was applied to a population it did not certify.
Within-frame calibration metrics could not have flagged the failure,
because they were computed within the frame
\citep{Steyerberg2009,Riley2019}.

\section{The meta-level no-free-lunch}
\label{sec:nfl}

At this point, a methodologically inclined reader has a natural reply:
\emph{very well, frames are real and choices are upstream of inference ---
let us formalise the choice. Develop a meta-procedure for deciding when a
frame is adequate; apply it. The frame problem dissolves into a
better-engineered second-order procedure.} The move is intellectually
attractive. It is also exactly the move the structural argument
forecloses.

Any meta-procedure for deciding frame adequacy is itself a procedure. As
such, it operates on a finite specification: of what to check, of what
counts as adequate, of what the test class includes. It is itself a frame
and inherits the original problem one level up. The regress is not
contingent on cleverness; it is structural. Choosing a formal apparatus
is itself a choice, and choosing the choice of formal apparatus is a
meta-choice with the same form.

\citet{WolpertMacready1997} formalised a particular version for
optimization: no algorithm is uniformly better than another across all
possible objective functions; in a fully-uninformed average, all
algorithms perform equally. \citet{Wolpert1996} stated an analogous
result for supervised learning. The no-free-lunch theorems are sometimes
received as technicalities that vanish under reasonable assumptions about
the function class. That reading is correct for the specific theorems and
misses the structural lesson, which generalises beyond the theorems'
formal scope. The lesson is that universal procedures forgo their
universality the moment they impose a frame --- and any practically useful
procedure must impose one --- so the frame is the choice being made.

The philosophical version is older. \citet{Hume1748} argued that any
inductive inference rests on a uniformity assumption that cannot itself
be established inductively without circularity. Hume's regress is the
philosophical antecedent of the meta-level argument: the assumption
that the future will resemble the past is upstream of induction and is
not certified by induction. \citet{Goodman1955} sharpened the point ---
the \emph{respects} in which the future is taken to resemble the past
are themselves an upstream choice, and induction does not certify that
choice either. The contemporary statistical traditions that have
grappled most rigorously with this include the subjectivist Bayesian
foundations \citep{deFinetti1974,Savage1954,Jaynes2003} and the
robust-Bayesian literature \citep{Berger1985,RiosInsuaRuggeri2000}; each
acknowledges, in its own register, that some choices are upstream of the
formalism and remain choices.

It is important to be clear about what the meta-level no-free-lunch is and
is not. It is not a claim that formal tools are useless: they are
indispensable. It is not a claim that any frame is as good as any other:
some frames are demonstrably better for some problems. It is a claim that
no formal apparatus delivers \emph{frame-choice} as a derived result. The
choice is upstream, and it stays upstream regardless of how many levels
of formalisation the discipline introduces.

The practical implication is not that quantification is hopeless. It is
that ``we just need better methodology'' is not an answer to the
structural problem. Better methodology is the choice being deferred. The
discipline's natural response to applied critique --- to formalise
further --- relocates the problem; it does not dissolve it. What remains
is dispositional work: how the practitioner, the organisation, and the
discipline hold the gap that no methodology closes.

\section{Automation and the erosion of reflexivity}
\label{sec:automation}

It is worth keeping two layers of the argument explicitly distinct. The
\emph{structural} layer (Sections~\ref{sec:picture}--\ref{sec:nfl}) is
permanent: it would have applied to nineteenth-century quantitative
practice and will apply to whatever practice supersedes the current one.
The
\emph{conjunctural} layer, developed in this section, is contingent: it
concerns the specific operational conditions of automated inference at
scale, and is in principle reshapable by institutional and technological
change. Conflating the two layers invites two equal and opposite
dismissals --- \emph{``models will get better, this will pass''} addresses
the conjunctural but ignores the structural; \emph{``these are eternal
problems, why worry now''} acknowledges the structural but misses the
conjunctural sharpening. The argument requires both: the structural
feature is what makes the conjunctural change matter, and the
conjunctural change is what makes the structural feature operationally
urgent.

For most of the history of quantitative practice, information processing
was slow, effortful, and human-mediated. The analyst who spent weeks
cleaning data, selecting features, fitting models, and writing up
conclusions was also performing a continuous act of epistemic scrutiny.
Assumptions were surfaced by friction; the frame was visible because
someone had to construct it, piece by piece, and could be challenged at
each step. The methodological conventions of twentieth-century statistics
--- model criticism, diagnostic residual analysis, sensitivity to prior
choice, careful reporting of conditions --- were institutionalised
expressions of that friction
\citep{Box1976,TukeyEDA1977,Cox2006,GelmanShalizi2013}.

Automation removes much of that friction. This is largely a gain: speed,
scale, reproducibility, accessibility. But the friction had a function.
It sustained a reflective loop in which a human held the uncertainty,
interrogated it, and decided how much confidence the output warranted.
That loop has not been replaced by automation; it has been bypassed.
Three features of the current moment compound the bypass, and they
reinforce one another.

The first is the opacity of relevance choices. When a human analyst
selects features, the selection is visible and arguable; a colleague can
ask, ``why this variable and not that one?'' When a deep neural network
learns its own representation from raw data, the same relevance choices
are still being made, but they are now embedded in architecture, training
data, and optimisation, no longer legible or easily contestable
\citep{Burrell2016,DoshiVelezKim2017}. The frame did not disappear; it
became invisible. The rhetoric of ``model-free'' methods, extended beyond
its technical scope, can encourage the belief that the inductive bias of
the learned model is absent rather than relocated. It is relocated ---
into the training distribution, the loss function, the architecture, and
the deployment integration --- and the relocation is the thing that needs
naming.

A second feature follows from the first, in a different register. Tools
that once produced predictions slowly and effortfully now produce them at
volumes that outpace any plausible rate of evaluation. The bottleneck
has flipped: it is no longer hard to generate answers; it is hard to
know which answers are reliable \citep{Bender2021}. This is especially
visible in the case of large language models, which produce fluent,
detailed, authoritative-sounding output with no built-in mechanism for
flagging when they are operating outside their competence. Fluency is
not knowledge, but at sufficient fluency the consumer who does not
already possess the underlying knowledge has no reliable way to detect
the divergence.

The third feature is a resource-allocation pattern that follows from the
first two and that is worth stating explicitly, since it is the
contemporary institutional version of the structural argument made
operational. As tools for producing predictions become cheaper and more
accessible, the scarce resource becomes the capacity to evaluate those
predictions --- domain knowledge, statistical judgment, structural
scepticism, an understanding of where models break. Expert judgment
becomes \emph{more} necessary precisely as institutions invest \emph{less}
in it, redirecting resources toward the production tools themselves. The
discipline of evaluation runs against the throughput incentives of
automated systems, and absent active protection, it will be hollowed out
in exactly the places where it is needed.

A related observation about contemporary practice deserves naming. \citet{Goodhart1975} and
\citet{Strathern1997} described how a measure, when made a target, ceases
to be a good measure. The applied implication for machine learning is
direct: when an evaluation metric becomes the optimisation target,
within-frame performance on the metric improves while the underlying
construct the metric was intended to capture drifts away. Benchmark
saturation, reward hacking, specification gaming, and distribution shift
are not unrelated phenomena; they are different operational lenses on the
same structural fact \citep{Amodei2016,Christian2020}. The frame was
fixed, and the world moved.

Confident outputs at scale, paired with the language of agency in which
modern AI systems are routinely described, encourage a perceptual drift:
from \emph{``a very complex tool operating within a frame''} to
\emph{``an intelligent agent whose capabilities might exceed the limits
that apply to all reasoning.''} They do not. The frame problem does not
close at sufficient compute; it is obscured by it. What was effortful and
visible in human inference is, in automated inference, fast and
invisible \citep{RussellS2019,Christian2020}. The structural limit is
unchanged; only its operational salience has shifted.

\section{The argument across practices}
\label{sec:practices}

A structural argument earns its keep by mattering for particular
practices. This section walks the argument through five figures of
contemporary quantitative work, identifying each figure's legitimate
entry point, the natural defense each brings, and the structural
counter-move. The figures are stylised --- no individual practitioner is
purely any of them --- but they capture five clusters of legitimate
professional commitment that the argument must engage on its own terms if
it is to be received as addressed to working practice rather than to
none.

\subsection{The engineer}
\label{sec:engineer}

The engineer's legitimate entry point is decisions with real
consequences: a built thing that performs or fails, a system in
deployment, a margin between safe and not-safe. The engineer grants
immediately that predictions are imperfect and that systems surprise
their designers. The natural question is what the practiced methodology
of safety factors, redundancy, and conservative design buys, and where it
does not yet handle the issue. Engineering has spent a century
institutionalising the acknowledgment that analytical models do not cover
everything; the safety factor is the cultural inheritance of many
failures \citep{Petroski1985,Perrow1984,Vaughan1996}.

The engineer's natural defense is that frame inadequacy is already priced
into practice. The defense is partly right --- which is why it works.
The counter-move is not to deny the heuristics but to show where they
predictably fail. Safety factors address \emph{anticipated} failure modes.
A multiplier of $1.5$ on a structural load assumes that the failure modes
on the analyst's list are the relevant ones, and that the residue is
quantitatively similar to what is on the list. Both assumptions are
within-frame. The failure that was never on the list --- the coupling at
the interface between two models, the regime outside calibration, the
variable excluded before the analysis began --- is the one safety factors
are structurally silent about.

Engineering practice supplies a concrete and recurrent illustration of
how within-frame rigor can accumulate to global overconfidence, in what
we will call the \emph{chained-model pattern}. Complex systems --- in
aerospace, energy, climate, medicine --- are too complicated for any
single model to span. The discipline divides: a thermal expert models
heat transfer, a structural expert models mechanical loads, a materials
expert characterises properties. Each is careful within their scope. The
outputs are then chained, and each link treats the previous link's
output as a given --- a number, an input, a fact --- not as what it
actually is, which is a prediction made under assumptions and carrying
uncertainty that has now become invisible. Each interface converts a
conditional output into an unconditional input, laundering a layer of
uncertainty. By the time the full chain is assembled, the result has
clean convergence, eight significant digits, and a stack of within-frame
calibrations sitting under it that no one propagated
\citep{OberkampfRoy2010,RoyOberkampf2011}.

The engineer's practice-implicating exposure is therefore that
uncertainty quantification typically distinguishes aleatory and epistemic
and rarely raises the ontological category. The practical move that
follows is to widen the distribution of considered outcomes deliberately:
to include a stress case, an alternative dreary scenario, an ``and what
if we are missing something'' branch. The move resists formal
specification; it cannot be derived from inside the frame. It is what
epistemic humility looks like in engineering practice when consequences
land. The communication move
is parallel: a number reported without its conditions, assumptions, and
failure modes is not a transfer of information but a transfer of
responsibility dressed as information.

\subsection{The statistician}
\label{sec:statistician}

The statistician's legitimate entry point is the proper use of tools ---
probability, inference, modelling, uncertainty quantification --- as a
craft with right and wrong ways to deploy each piece of machinery. The
natural question is what conditions the methods produce reliable
conclusions under, and where those conditions are silently violated in
deployment. The statistician is unusually well-equipped to follow the
structural argument because the argument is, in part, a recapitulation of
the discipline's own foundations \citep{Cox2006,Lindley2006}.

A foundational point load-bears here, and it is worth stating early.
Probability is most usefully read as a support tool for judgment under
background knowledge, distributing plausibility over possibilities; it is
not, in this reading, a quantity grounded in nature
\citep{deFinetti1974,Savage1954,Jaynes2003,Lindley2006}. The mathematics
is identical to a frequentist reading whenever probability can be
identified with the relative frequency of a repeatable event whose
reference class is given by the physical setup --- coin flips, items
off an assembly line. The interpretive difference matters when no such
natural reference class is available, as with one-off events or
situations in which the population of comparable cases has to be drawn
by judgment rather than counted from the world. In those cases the
construction of the reference class is itself a frame choice, and the
formalism does not certify it. The framing of
``distributing over possible worlds given background knowledge'' is not
a license for arbitrary belief --- it is a description of where the
inference is anchored, and the anchoring is upstream of the formalism.
The common framing that the Bayesian commitment is ``subjective'' is
correct; the inference that ``subjective'' means ``unprincipled'' is a
mistake. Subjectivity here names a structural feature of inference under
finite information.

The statistician's natural defense is that the methodology is sound and
that deployment failures are misapplications rather than features of the
methods. The counter-move is not to deny that the methodology is sound
but to show that some deployment failures are not misapplications. They
are structural consequences of the tool's design conditions being
silently relaxed at the institutional interface. Robustness analysis is
bounded by an assumed model class. Sensitivity analysis perturbs
parameters within a frame; it does not relax the frame. Calibration is
computed against a reference class that is itself a choice. Each of
these tools is exactly what it advertises. Collectively, they
underdetermine adequacy of the frame, and they do so structurally.

The statistician is exposed in particular to two practice-implicating
claims. First, ontological uncertainty --- what is not in the model
because no one thought to put it there --- is the category the
discipline's apparatus most underweights; adjacent traditions ---
notably risk analysis, policy modelling under uncertainty, and the
post-normal science framework --- sometimes name this category
explicitly where statistics-as-such does not
\citep{Wynne1992,Walker2003,FuntowiczRavetz1990}.
Second, bounded robustness is not the same object as indeterminate
framing: the residue that sensitivity analysis bounds lives inside the
frame, while some of the practically consequential residue is the frame,
not in it. \citet{GelmanShalizi2013}, working within the practice of
Bayesian statistics, argue for a model-checking-as-frame-revision posture
that is consistent with the structural claim --- model checking is part
of how new relevance enters the inference, and it is dispositional rather
than reducible to a closed procedure.

The practical landing for the statistician is a craft commitment: report
predictive uncertainty in a way that distinguishes what the frame bounds
from what the frame is; propagate uncertainty rather than launder it at
every interface; name conditions rather than transferring responsibility
through a clean number. Not stylistic preference --- the rigor of knowing
which rigor was earned.

\subsection{The mathematician}
\label{sec:mathematician}

The mathematician's legitimate entry point is formal objects and their
structural properties: definitions, theorems, limits, impossibility
results. The natural question is what the structural reason is ---
independent of any particular tool's adequacy --- for incomplete
reasoning systems, and what that reason implies for what can be proved.

The mathematician's natural defense is the clean one: the formal object
does what it does; whether it is the right tool for a given purpose is a
separate question and not a critique of the object. The defense is
correct on its own terms. The structural counter-move is not to attack it
but to push the argument up the abstraction ladder. The claim is not that
any particular formal tool fails its job. The claim is that the act of
choosing which formal tool to apply is itself a frame-dependent choice,
and no formal system can decide it for you from inside. This is the
meta-level no-free-lunch (Section~\ref{sec:nfl}) in its sharpest form.

For the mathematician, the structural argument lands most directly. The
arc runs through the formal frame problem (relevance filtering, finite
specification, the impossibility of self-audit), through the
deduction--induction distinction in which both modes inherit the frame
problem and deduction does not escape it through formality, and arrives
at probability and Bayesian decision theory as the most refined formal
response to the structural problem. The structural problem then reappears
at the meta-level in the form of prior selection, utility specification,
and the choice of reference class.

The discipline's natural response to applied critique is to formalise
further. The structural argument says: formalising further is not the
closure of the problem but its relocation one level up. This is not a
counsel against formalisation --- formalisation is the discipline's
contribution. It is a counsel against the rhetorical move that treats
formalisation as a solvent. The solvent is not in the formalism; the
formalism is the next thing the structural problem applies to.

\subsection{The machine-learning practitioner}
\label{sec:mlpractitioner}

The machine-learning practitioner's legitimate entry point is what the
system actually does: empirical performance on tasks, deployed behavior,
capability boundaries, what scales and what does not. The natural
question is what the system inherits, and where its capabilities end in
ways the benchmarks do not say.

The natural defense is the most distinctive of the five, and it deserves
direct address: \emph{the frame problem is old-AI thinking; modern ML is
a different paradigm; learned representations route around it}. This is
the load-bearing resistance, and the argument fails for this audience if
it does not engage it in modern idiom.

The first move it has to make is translation. The frame did not
disappear when the practice stopped writing symbolic rules; it moved.
What earlier generations encoded in hand-written rules is now encoded in
the training distribution, the loss function, the architecture, the
data curation, and the deployment integration. Every learned predictor
works on a finite specification of conditions --- the conditions it was
trained on --- and what falls outside that specification does not appear
as wider uncertainty bars. \emph{Distribution shift} is the operational
name for exactly this \citep{QuioneroCandela2009}; \emph{reward hacking}
is the operational name for cases in which the system's objective
specification is a frame that does not match the construct the designer
intended; \emph{specification gaming} is the same fact in slightly
different operational dress \citep{Amodei2016,Christian2020}. The recent
literature on Bayesian deep learning has begun to address the
aleatory/epistemic distinction inside modern ML
\citep{KendallGal2017,HullermeierWaegeman2021}, and this work is valuable
in that it extends within-frame uncertainty quantification to deep
models. But it is, by construction, within-frame uncertainty
quantification: the ontological residue --- the part that comes from the
choice of training distribution, loss function, and architecture being
themselves a frame --- remains outside the formal apparatus, as it does
for any UQ inside an assumed model class.

The second move is to address scale. The practitioner's working
hypothesis --- that more data, more parameters, and more compute will
typically solve more problems --- is partly right, which is exactly why
it works as a defense. What is not right is the implication that scale
dissolves the underlying limit. Pattern-fit at scale is still pattern-fit
inside a frame, and generality with respect to a training distribution
is not the same object as generality with respect to the world. The
system reports calibrated confidence over the things it was trained to
consider; it has no representation of the things it was not. Scale does
not dissolve the frame problem; it obscures it \citep{Bender2021}.

The third move is to address opacity. The abstraction ladder underneath
modern ML --- silicon, operating systems, frameworks, learned
representations --- is a remarkable feat of accumulated engineering, but
its very elaborateness renders the system's inner workings opaque, even
to trained practitioners \citep{Burrell2016,DoshiVelezKim2017}. Opacity
is not neutral: it makes the rhetorical drift from \emph{``a complex tool
operating within a frame''} to \emph{``an intelligent agent whose
capabilities might exceed the limits that apply to all reasoning''}
easier rather than harder to slide into. Recognising this drift as a
perceptual error rather than as a statement about capability is one of
the practical loads the frame argument carries for this audience.

What this asks of practice is specific. Hold the system's confidence as
a property of the system, not of the world. Treat the observation that a
system was confident as different information from the observation that
the system was correct. Distinguish out-of-distribution detection ---
which is a within-frame proxy --- from adequacy with respect to the
world, to which the system has no access. Recognise evaluation as a
scarce resource that needs to be protected against the throughput
pressure of the production pipeline, rather than as an unlimited one
that can be assumed to scale with the production tools.

\subsection{The non-specialist recipient of expert claims}
\label{sec:layperson}

The non-specialist's legitimate entry point is concrete situations with
stakes that can be held without specialist machinery --- a decision, a
choice, a story. The non-specialist does not have a ``practice'' in the
sense the four professionals do; the bearing of the frame argument is
not on destabilising a practice but on equipping a stance with which to
receive expert claims that are being made \emph{to} them.

What distinguishes this audience from the previous four is an asymmetry.
The non-specialist bears the consequences of a framing decision they did
not make and could not, from their position, have audited. The
professional figures inhabit a practice that frames its own objects; the
non-specialist inhabits the \emph{output} of a practice they do not see.
The frame argument bears on them as the structural reason a confident
expert output is not what it looks like, and the disposition it earns is
one of intelligent reception rather than internal practice.

The stance has a few specific contents that hang together. The first is
that when an expert gives a number, the number is conditional on a frame
the expert chose; asking what that frame is, and what was left out of
it, is not a failure to understand the technical material but is the
substantive content of intelligent reception of expert advice. The
non-specialist does not need to evaluate the technical work; they need
to know that the \emph{conditions} of the work are part of what was
communicated, and they need to ask for them when they are not. The
second is that confidence and reliability are different things: a system
that produces confident outputs faster than they can be evaluated is not
the same as a system that produces reliable outputs, even if the two are
hard to tell apart from outside. The third is that expressed uncertainty
is not always a failure of expertise; sometimes the absence of expressed
uncertainty is the warning sign rather than the evidence of competence.
The fourth is that some questions are not the kind of question that has
an answer in the form being asked, and recognising that is itself useful
knowledge.

These contents convert into a small set of legible questions the
non-specialist can put to any confident output, regardless of domain:
\emph{under what conditions does this hold, what would change the
number, what is the most likely way this is wrong, and what was
deliberately not modelled?} The questions do not require technical
competence; they require the disposition to ask them, and the
willingness to receive ``I don't know'' or ``we did not consider that''
as substantive answers rather than as failures of expertise. A
respondent who cannot or will not engage these questions has, in effect,
asked the non-specialist to absorb a frame as if it were a fact --- the
transfer of responsibility dressed as information named earlier
(Section~\ref{sec:engineer}) reappears here as the structural risk the
non-specialist actually faces.

The natural defense the non-specialist brings to these claims is rarely
ideological; it is attentional --- ``this has left the world of things I
can picture; I am not the audience for this.'' The counter-move is not
argument; it is keeping a concrete anchor in view long enough for the
abstraction to settle into recognisable form. The anchor is the stakes:
the non-specialist knows what is at stake for them in a way no expert
frame can guarantee to have captured, and that knowledge --- irreducible
to the expert's machinery --- is the legitimate ground from which
framing questions can be asked.

The contemporary situation sharpens this exposure rather than easing it.
Automated systems increasingly produce confident expert outputs without
any single human expert at the other end whose framing choices could
even in principle be interrogated. A model output, a generated summary,
a personalised recommendation --- these are received by non-specialists
as expert claims, but the frame is now doubly removed: it sits in
training data, architecture, and deployment integration, and is not
legible to the producer of the output, let alone to its recipient. The
disposition the frame argument earns becomes correspondingly more
important: the questions to ask of an output stay the same, but the
entity that must answer them is now an institution rather than a person,
and the institutional capacity to answer them is, by the argument of
Section~\ref{sec:automation}, exactly what is being hollowed out by the
throughput pressures of automated production.

The weather forecast illustrates the positive form of this stance. The
forecast that
says ``it will rain tomorrow'' is a confident prediction; the forecast
that says ``70\% chance of rain'' is more useful, not less. The first
lets the recipient bring an umbrella or not. The second supports a
decision proportional to stakes: how much getting wet matters, how
costly a wrong call is, what alternatives are available
\citep{Spiegelhalter2019,GneitingRaftery2007,TetlockGardner2015}. The
difference between a prediction and a decision tool is exactly what the
present argument makes visible across all the audiences it addresses
--- and to the non-specialist most of all, because for them the
difference between the two is whether they have been given information
or have been issued a directive in the costume of information.

\section{Epistemic humility as structural commitment}
\label{sec:humility}

The argument developed across the preceding sections --- that reasoning
runs on a frame chosen upstream of inference, that the frame is
structurally invisible from inside, that the residue of the frame
(``ontological uncertainty'') is not addressable from within it, and
that no meta-procedure dissolves the regress --- earns a particular
practical disposition. The disposition is epistemic humility: not as a
stylistic preference, not as a moral one, but as the practical posture
consonant with what the argument shows.

The disposition has, in practice, four specific commitments, each of
which follows from a claim already developed. The first is to hold
belief provisionally, in proportion to the conditions under which the
inference was made and the confidence those conditions warrant; this
follows from the fact that the frame is finite and that the residue of
the frame is invisible from within it. The second is to propagate
uncertainty between stages of a chained inference rather than launder it
at each handoff; this follows from the observation, made concrete in the
chained-model pattern of Section~\ref{sec:engineer}, that the
within-frame calibration of each link does not certify the chain. The
third is to communicate conditions alongside conclusions, so the
recipient of an analysis knows what they are receiving; the reason is
that a prediction issued without its conditions delegates to the
consumer a task --- inferring the conditions --- that the consumer is
typically not in a position to perform. The fourth is to protect
evaluative capacity actively, in institutional and pedagogical practice,
because automation accelerates the production of confident output faster
than evaluative resources accumulate, and the asymmetry will not correct
itself.

It is important to name what this disposition is not. It is not
paralysis: the argument does not recommend that practitioners stop
making predictions, building systems, or offering judgments. It
recommends, rather, that they make predictions, build systems, and offer
judgments with the conditions of those acts visible. Nor is it false
modesty. The disposition is not a performance of uncertainty designed to
forestall criticism; it is a substantive commitment to a particular kind
of rigor --- the rigor of knowing which rigor was earned, and which
assumptions are silently relaxed at the interface where the model meets
the world.

The disposition is costly. Structural incentives reward confident
outputs over honestly conditional ones
\citep{IoannidisPLoS2005,NosekOSC2015}. Academic publishing favours
definitive findings; product development favours shipped features;
consulting favours confident recommendations. Psychological commitments
to expertise resist provisional belief; the practitioner who says ``this
is probably right, under these conditions which may not hold'' rarely
receives the contract, the funding, or the airtime. Scientific training itself
rewards the discovery and confident assertion of correct results, not
calibrated uncertainty or the practice of holding conclusions
provisionally; \citet{Tetlock2005}, in a longitudinal study of expert
forecasters, documented the typical consequence: experts tend to
overconfidence, and the more confident among them tend to be less
accurate; \citet{Kahneman2011} synthesises the broader cognitive
psychology of the same pattern. These costs are real and need to be
named honestly;
pretending they do not exist is itself a frame failure.

The costs of the disposition need to be set against the costs of the
alternative. The historical record makes the comparison concrete: the
Challenger disaster, the 2008 financial crisis, the replication crisis
in psychology and biomedical sciences, and recent deployment failures in
automated decision systems and clinical prediction models. In each, the
failure was not a lack of information. It was the systematic suppression
of expressed uncertainty by institutions that could not --- technically,
organisationally, or psychologically --- tolerate it. The cost of that
suppression, when it came due, was orders of magnitude greater than what
honest uncertainty would have imposed along the way
\citep{Vaughan1996,IoannidisPLoS2005,KingKay2020}.

Epistemic humility does not ask for more doubt. It asks for doubt in the
right places, communicated in the right form, with the rigor of knowing
which doubts are structural and which are within-frame.

The formulation often attributed to Keynes --- \emph{``it is better to
be vaguely right than precisely wrong''} --- is in fact older
\citep{Read1898}; the attribution is folkloric. The maxim is sharper
than it is usually heard to be. It is not a counsel of vagueness. It is
a counsel for a different kind of rigor: of knowing where precision was
earned and where it was assumed, of propagating that awareness through
the chain of inference rather than laundering it, and of designing for
robustness to surprise rather than optimising for the anticipated case.
The disposition does not improve with scale, does not yield to
automation, and cannot be replaced by methodology. It is the remainder.

\section{Implications for practice, organisation, and pedagogy}
\label{sec:discussion}

The implications of the foregoing distribute across four levels of
activity: the people who build models, the people who decide on their
basis, the institutions that house both, and the curricula that train
the next generation. We sketch each briefly; the aim is to indicate the
shape of the practical move rather than to legislate specific
procedures.

For those who build models, the most consequential decisions in any
modelling exercise are made before training or fitting begins: what to
measure, what to include, what to treat as fixed, what to ignore. These
decisions structurally underdetermine model adequacy and structurally
receive less scrutiny than architecture choice or hyperparameter tuning,
partly because they are harder to formalise and partly because they are
made early and then forgotten. A useful diagnostic, applied to any
consequential prediction the team produces, is to ask what the output
would look like if one upstream assumption were different; to ask which
assumptions cannot be tested even in principle; and to ask whether the
current uncertainty quantification would have flagged the failure modes
one can imagine. If the answer to the last question is ``probably
not,'' that may be the most important finding of the analysis.
Propagation methods are widely available and well understood; what is
typically absent is the discipline to use them, because propagated
uncertainty is always wider and less comfortable than the laundered
version.

For decision-makers who consume model outputs, the institutional
communication chain strips uncertainty at every interface, so that the
person at the end of the chain, who allocates resources and bears
consequences, typically sees only the final number. In the formulation
of \citet{Spiegelhalter2019}, a number that comes with no account of its
conditions, assumptions, and failure modes is not a decision tool. A
prediction of $X$ is meaningful only once one knows $X$ under what
conditions, what was assumed, what would change the number, and what
the most likely structural cause of a deployment failure would be. If
the team producing the prediction cannot answer these questions
concisely, the prediction is not ready to support a consequential
decision, regardless of how precise it appears.

At the institutional level, the practical claim is that every
investment in systems that generate confident output should be paired
with investment in the capacity to evaluate that output. If production
is automated and evaluation is hollowed out, the result is a factory at
full speed with no quality control. This means protecting domain
expertise as a resource rather than a cost, funding adversarial review
processes that do not need to justify themselves on a per-engagement
basis, creating organisational spaces in which honest uncertainty is
rewarded rather than punished, and recognising that the discipline of
evaluation runs against the throughput incentives of automated systems
and will therefore not survive on those incentives alone. The
disposition described in Section~\ref{sec:humility} cannot be sustained
by individual practitioners; it has to be supported by organisational
design \citep{Vaughan1996,KingKay2020}.

For pedagogy, the same logic applies one level upstream. Teaching
quantitative practice without teaching the frame problem produces a
practitioner who is competent within frames they did not choose. The
foundational distinction between within-frame rigor and adequacy of the
frame deserves a place early in the curriculum, not as a philosophical
prelude to be hurried past on the way to technique, but as the
structural setting that determines what technique can and cannot do.
\citet{Box1976}, \citet{Knight1921}, \citet{Jaynes2003}, \citet{Pearl1988},
\citet{WolpertMacready1997}, and \citet{Spiegelhalter2019} belong in
early methodological education for the same reason --- not for their
conclusions on specific problems but for the disposition their work
documents and trains. That disposition is harder to teach than
technique; it is also less reducible to assessment, which is one of the
reasons it tends to be deferred. The argument of this paper is that the
deferral is not optional.

A further observation about practice belongs here, because it answers
one objection a careful reader will already have anticipated. If a
reasoning system cannot audit its own frame from within, the question
of how new relevance ever enters the system becomes important. There
are essentially two routes by which it can. Either someone tells the
agent --- another vantage, another discipline, another witness who is
not inside the same frame --- or the agent bumps into the new relevance
through exploration of the world
\citep{SuttonBarto2018,RussellS2019}. Exploration is therefore not
optional in this picture; it is one of the few mechanisms by which the
frame is updatable from outside. It is also not free. It does not
escape the finite-specification issue, because choices about what to
explore are themselves frame-conditional. It has no built-in guarantee
of discovery. And it is practically limited by stakes, because some
explorations are unacceptable on grounds of damage, risk, or
irreversibility. Acknowledging this is part of the disposition the
argument earns: exploration is one of the partial answers to the
structural problem, and the conditions under which exploration is
possible and responsible are themselves matters for institutional and
technical design.

Put crudely, as a practical maxim that follows from this section:
explore where exploration is cheap, reversible, and supported by some
intuition for what one might learn; stay receptive to external critique
and disposed to self-critique, since being told from outside the frame
is the other route by which new relevance enters; and admit an
uncertainty buffer --- a deliberate widening of the distribution of
considered outcomes --- where stakes are high, actions are
irreversible, or the prediction map is fragile to small changes in the
frame. Again, this is not the formalisation of a solution --- any such
formalisation should be regarded with caution, by the meta-level
argument of Section~\ref{sec:nfl} --- but the naming of a
context-sensitive practical posture: flexibility and adaptability where
formal methods cannot reach closure.

\section{Conclusion}
\label{sec:conclusion}

We act on the world through pictures we make of it. The pictures are
always partial because the picture has to be specified finitely from
somewhere, and somewhere is not nowhere. The world responds to the
action, not to the picture. Closing the gap between the two is the basic
structure of any rational endeavor; understanding why the gap cannot be
closed completely is the basic structure of an intellectually honest
quantitative practice.

The structural argument runs through nine connected claims:
\emph{decisions carry stakes; reasoning runs on a frame whose relevance
must be filtered before any inference can run; the specification of that
frame is necessarily finite; three categories of uncertainty operate, the
third of which is structurally invisible from within the frame; the limit
applies to deduction as much as to induction; the reasoning system cannot
audit its own frame from within; no meta-procedure dissolves the regress; 
epistemic humility is the disposition the argument earns;
automation now raises the cost of failing to hold the disposition.}
The argument is not against rigor. It
is for a particular kind of rigor: the rigor of knowing what your
precision was earned over and what it was assumed over; of propagating
that awareness through the chain of inference rather than laundering it;
of designing for robustness to surprise rather than optimising for the
anticipated case; and of protecting the human and organisational
capacity to inspect frames, because automation does not preserve that
capacity by default.

That discipline has to be actively cultivated --- before the pace of
automation erodes the reflective capacity that makes it possible, and
makes us, as it will on some consequential issue, \emph{precisely wrong}.



\begin{thebibliography}{0}
\providecommand{\natexlab}[1]{#1}
\providecommand{\url}[1]{\texttt{#1}}
\expandafter\ifx\csname urlstyle\endcsname\relax
  \providecommand{\doi}[1]{doi: #1}\else
  \providecommand{\doi}{doi: \begingroup \urlstyle{rm}\Url}\fi

\end{thebibliography}


\begin{thebibliography}{99}
\setlength{\itemsep}{0.1em}

\bibitem[Amodei et~al.(2016)]{Amodei2016}
Amodei, D., Olah, C., Steinhardt, J., Christiano, P., Schulman, J., \&
Man\'e, D. (2016). Concrete problems in AI safety. \emph{arXiv preprint}
arXiv:1606.06565.

\bibitem[Angelopoulos \& Bates(2023)]{AngelopoulosBates2023}
Angelopoulos, A.~N., \& Bates, S. (2023). Conformal prediction: A
gentle introduction. \emph{Foundations and Trends in Machine Learning},
16(4), 494--591.

\bibitem[Bender et~al.(2021)]{Bender2021}
Bender, E.~M., Gebru, T., McMillan-Major, A., \& Shmitchell, S. (2021).
On the dangers of stochastic parrots: Can language models be too big?
\emph{Proceedings of the 2021 ACM Conference on Fairness, Accountability,
and Transparency}, 610--623.

\bibitem[Berger(1985)]{Berger1985}
Berger, J.~O. (1985). \emph{Statistical Decision Theory and Bayesian
Analysis} (2nd ed.). New York: Springer.

\bibitem[Box(1976)]{Box1976}
Box, G.~E.~P. (1976). Science and statistics. \emph{Journal of the
American Statistical Association}, 71(356), 791--799.

\bibitem[Box(1979)]{Box1979}
Box, G.~E.~P. (1979). Robustness in the strategy of scientific model
building. In R.~L. Launer \& G.~N. Wilkinson (Eds.), \emph{Robustness in
Statistics} (pp.~201--236). New York: Academic Press.

\bibitem[Burrell(2016)]{Burrell2016}
Burrell, J. (2016). How the machine `thinks': Understanding opacity in
machine learning algorithms. \emph{Big Data \& Society}, 3(1).

\bibitem[Christian(2020)]{Christian2020}
Christian, B. (2020). \emph{The Alignment Problem: Machine Learning and
Human Values}. New York: W.~W. Norton.

\bibitem[Cox(2006)]{Cox2006}
Cox, D.~R. (2006). \emph{Principles of Statistical Inference}.
Cambridge: Cambridge University Press.

\bibitem[de Finetti(1974)]{deFinetti1974}
de Finetti, B. (1974). \emph{Theory of Probability} (Vols.~1--2). New
York: Wiley.

\bibitem[Dennett(1984)]{Dennett1984}
Dennett, D.~C. (1984). Cognitive wheels: The frame problem of AI. In
C.~Hookway (Ed.), \emph{Minds, Machines and Evolution} (pp.~129--150).
Cambridge: Cambridge University Press.

\bibitem[Doshi-Velez \& Kim(2017)]{DoshiVelezKim2017}
Doshi-Velez, F., \& Kim, B. (2017). Towards a rigorous science of
interpretable machine learning. \emph{arXiv preprint} arXiv:1702.08608.

\bibitem[Draper(1995)]{DraperJRSS1995}
Draper, D. (1995). Assessment and propagation of model uncertainty.
\emph{Journal of the Royal Statistical Society: Series B}, 57(1), 45--97.

\bibitem[Fauriat(2022)]{Fauriat2022}
Fauriat, W. (2022). \emph{Un discours de la m\'ethode au XXIe si\`ecle:
Risques et limites de l'\guillemotleft\,intelligence artificielle\,\guillemotright}.
Auto-\'edition. ISBN 979-10-40-50265-4.

\bibitem[Funtowicz \& Ravetz(1990)]{FuntowiczRavetz1990}
Funtowicz, S.~O., \& Ravetz, J.~R. (1990). \emph{Uncertainty and Quality
in Science for Policy}. Dordrecht: Kluwer.

\bibitem[Gelman \& Shalizi(2013)]{GelmanShalizi2013}
Gelman, A., \& Shalizi, C.~R. (2013). Philosophy and the practice of
Bayesian statistics. \emph{British Journal of Mathematical and
Statistical Psychology}, 66(1), 8--38.

\bibitem[Gneiting \& Raftery(2007)]{GneitingRaftery2007}
Gneiting, T., \& Raftery, A.~E. (2007). Strictly proper scoring rules,
prediction, and estimation. \emph{Journal of the American Statistical
Association}, 102(477), 359--378.

\bibitem[Goodhart(1975)]{Goodhart1975}
Goodhart, C.~A.~E. (1975). Problems of monetary management: The U.K.
experience. \emph{Papers in Monetary Economics}, Reserve Bank of
Australia.

\bibitem[Goodman(1955)]{Goodman1955}
Goodman, N. (1955). \emph{Fact, Fiction, and Forecast}. Cambridge, MA:
Harvard University Press.

\bibitem[Helton \& Oberkampf(2004)]{HeltonOberkampf2004}
Helton, J.~C., \& Oberkampf, W.~L. (Eds.) (2004). Alternative
representations of epistemic uncertainty. Special issue,
\emph{Reliability Engineering \& System Safety}, 85(1--3).

\bibitem[Hern\'an \& Robins(2020)]{Hernan2020}
Hern\'an, M.~A., \& Robins, J.~M. (2020). \emph{Causal Inference: What
If}. Boca Raton: Chapman \& Hall/CRC.

\bibitem[Hoeting et~al.(1999)]{HoetingBMA1999}
Hoeting, J.~A., Madigan, D., Raftery, A.~E., \& Volinsky, C.~T. (1999).
Bayesian model averaging: A tutorial. \emph{Statistical Science}, 14(4),
382--417.

\bibitem[H\"ullermeier \& Waegeman(2021)]{HullermeierWaegeman2021}
H\"ullermeier, E., \& Waegeman, W. (2021). Aleatoric and epistemic
uncertainty in machine learning: An introduction to concepts and
methods. \emph{Machine Learning}, 110, 457--506.

\bibitem[Hume(1748)]{Hume1748}
Hume, D. (1748). \emph{An Enquiry Concerning Human Understanding}.
London: A.~Millar.

\bibitem[Ioannidis(2005)]{IoannidisPLoS2005}
Ioannidis, J.~P.~A. (2005). Why most published research findings are
false. \emph{PLoS Medicine}, 2(8), e124.

\bibitem[Jaynes(2003)]{Jaynes2003}
Jaynes, E.~T. (2003). \emph{Probability Theory: The Logic of Science}.
Cambridge: Cambridge University Press.

\bibitem[Kahneman(2011)]{Kahneman2011}
Kahneman, D. (2011). \emph{Thinking, Fast and Slow}. New York: Farrar,
Straus and Giroux.

\bibitem[Kendall \& Gal(2017)]{KendallGal2017}
Kendall, A., \& Gal, Y. (2017). What uncertainties do we need in Bayesian
deep learning for computer vision? \emph{Advances in Neural Information
Processing Systems}, 30, 5574--5584.

\bibitem[King \& Kay(2020)]{KingKay2020}
King, M., \& Kay, J. (2020). \emph{Radical Uncertainty: Decision-Making
Beyond the Numbers}. New York: W.~W. Norton.

\bibitem[Knight(1921)]{Knight1921}
Knight, F.~H. (1921). \emph{Risk, Uncertainty, and Profit}. Boston:
Houghton Mifflin.

\bibitem[Korzybski(1933)]{Korzybski1933}
Korzybski, A. (1933). \emph{Science and Sanity: An Introduction to
Non-Aristotelian Systems and General Semantics}. Lancaster, PA:
International Non-Aristotelian Library Publishing Company.

\bibitem[Lempert et~al.(2003)]{Lempert2003}
Lempert, R.~J., Popper, S.~W., \& Bankes, S.~C. (2003). \emph{Shaping
the Next One Hundred Years: New Methods for Quantitative, Long-Term
Policy Analysis}. Santa Monica: RAND.

\bibitem[Lindley(2006)]{Lindley2006}
Lindley, D.~V. (2006). \emph{Understanding Uncertainty}. Hoboken: Wiley.

\bibitem[McCarthy \& Hayes(1969)]{McCarthyHayes1969}
McCarthy, J., \& Hayes, P.~J. (1969). Some philosophical problems from
the standpoint of artificial intelligence. In B.~Meltzer \& D.~Michie
(Eds.), \emph{Machine Intelligence}, Vol.~4 (pp.~463--502). Edinburgh:
Edinburgh University Press.

\bibitem[Morgan \& Henrion(1990)]{MorganHenrion1990}
Morgan, M.~G., \& Henrion, M. (1990). \emph{Uncertainty: A Guide to
Dealing with Uncertainty in Quantitative Risk and Policy Analysis}.
Cambridge: Cambridge University Press.

\bibitem[Open Science Collaboration(2015)]{NosekOSC2015}
Open Science Collaboration (2015). Estimating the reproducibility of
psychological science. \emph{Science}, 349(6251), aac4716.

\bibitem[Oberkampf \& Roy(2010)]{OberkampfRoy2010}
Oberkampf, W.~L., \& Roy, C.~J. (2010). \emph{Verification and
Validation in Scientific Computing}. Cambridge: Cambridge University
Press.

\bibitem[Pearl(1988)]{Pearl1988}
Pearl, J. (1988). \emph{Probabilistic Reasoning in Intelligent Systems:
Networks of Plausible Inference}. San Mateo: Morgan Kaufmann.

\bibitem[Pearl(2009)]{Pearl2009}
Pearl, J. (2009). \emph{Causality: Models, Reasoning and Inference} (2nd
ed.). Cambridge: Cambridge University Press.

\bibitem[Perrow(1984)]{Perrow1984}
Perrow, C. (1984). \emph{Normal Accidents: Living with High-Risk
Technologies}. New York: Basic Books.

\bibitem[Petroski(1985)]{Petroski1985}
Petroski, H. (1985). \emph{To Engineer is Human: The Role of Failure in
Successful Design}. New York: St.~Martin's Press.

\bibitem[Pylyshyn(1987)]{Pylyshyn1987}
Pylyshyn, Z.~W. (Ed.) (1987). \emph{The Robot's Dilemma: The Frame
Problem in Artificial Intelligence}. Norwood: Ablex.

\bibitem[Qui\~nonero-Candela et~al.(2009)]{QuioneroCandela2009}
Qui\~nonero-Candela, J., Sugiyama, M., Schwaighofer, A., \& Lawrence,
N.~D. (Eds.) (2009). \emph{Dataset Shift in Machine Learning}.
Cambridge, MA: MIT Press.

\bibitem[Read(1898)]{Read1898}
Read, C. (1898). \emph{Logic, Deductive and Inductive}. London: Grant
Richards.

\bibitem[Riley et~al.(2019)]{Riley2019}
Riley, R.~D., van~der Windt, D., Croft, P., \& Moons, K.~G.~M.
(2019). \emph{Prognosis Research in Healthcare: Concepts, Methods, and
Impact}. Oxford: Oxford University Press.

\bibitem[R\'{\i}os Insua \& Ruggeri(2000)]{RiosInsuaRuggeri2000}
R\'{\i}os Insua, D., \& Ruggeri, F. (Eds.) (2000). \emph{Robust Bayesian
Analysis}. Lecture Notes in Statistics 152. New York: Springer.

\bibitem[Roy \& Oberkampf(2011)]{RoyOberkampf2011}
Roy, C.~J., \& Oberkampf, W.~L. (2011). A comprehensive framework for
verification, validation, and uncertainty quantification in scientific
computing. \emph{Computer Methods in Applied Mechanics and Engineering},
200(25--28), 2131--2144.

\bibitem[Russell(1912)]{Russell1912}
Russell, B. (1912). \emph{The Problems of Philosophy}. London: Williams
\& Norgate.

\bibitem[Russell(2019)]{RussellS2019}
Russell, S. (2019). \emph{Human Compatible: Artificial Intelligence and
the Problem of Control}. New York: Viking.

\bibitem[Saltelli et~al.(2008)]{Saltelli2008}
Saltelli, A., Ratto, M., Andres, T., Campolongo, F., Cariboni, J.,
Gatelli, D., Saisana, M., \& Tarantola, S. (2008). \emph{Global
Sensitivity Analysis: The Primer}. Chichester: Wiley.

\bibitem[Savage(1954)]{Savage1954}
Savage, L.~J. (1954). \emph{The Foundations of Statistics}. New York:
Wiley.

\bibitem[Shanahan(2016)]{Shanahan2016}
Shanahan, M. (2016). The frame problem. In E.~N.~Zalta (Ed.),
\emph{Stanford Encyclopedia of Philosophy} (Spring 2016 ed.).
Stanford: Metaphysics Research Lab.

\bibitem[Simon(1982)]{Simon1982}
Simon, H.~A. (1982). \emph{Models of Bounded Rationality} (Vols.~1--2).
Cambridge, MA: MIT Press.

\bibitem[Spiegelhalter(2019)]{Spiegelhalter2019}
Spiegelhalter, D. (2019). \emph{The Art of Statistics: Learning from
Data}. London: Penguin.

\bibitem[Steyerberg(2009)]{Steyerberg2009}
Steyerberg, E.~W. (2009). \emph{Clinical Prediction Models: A Practical
Approach to Development, Validation, and Updating}. New York: Springer.

\bibitem[Strathern(1997)]{Strathern1997}
Strathern, M. (1997). `Improving ratings': Audit in the British
University system. \emph{European Review}, 5(3), 305--321.

\bibitem[Sutton \& Barto(2018)]{SuttonBarto2018}
Sutton, R.~S., \& Barto, A.~G. (2018). \emph{Reinforcement Learning: An
Introduction} (2nd ed.). Cambridge, MA: MIT Press.

\bibitem[Taleb(2007)]{Taleb2007}
Taleb, N.~N. (2007). \emph{The Black Swan: The Impact of the Highly
Improbable}. New York: Random House.

\bibitem[Tetlock(2005)]{Tetlock2005}
Tetlock, P.~E. (2005). \emph{Expert Political Judgment: How Good Is It?
How Can We Know?}. Princeton: Princeton University Press.

\bibitem[Tetlock \& Gardner(2015)]{TetlockGardner2015}
Tetlock, P.~E., \& Gardner, D. (2015). \emph{Superforecasting: The Art
and Science of Prediction}. New York: Crown.

\bibitem[Tukey(1977)]{TukeyEDA1977}
Tukey, J.~W. (1977). \emph{Exploratory Data Analysis}. Reading, MA:
Addison-Wesley.

\bibitem[Vaughan(1996)]{Vaughan1996}
Vaughan, D. (1996). \emph{The Challenger Launch Decision: Risky
Technology, Culture, and Deviance at NASA}. Chicago: University of
Chicago Press.

\bibitem[Vovk et~al.(2005)]{Vovk2005}
Vovk, V., Gammerman, A., \& Shafer, G. (2005). \emph{Algorithmic
Learning in a Random World}. New York: Springer.

\bibitem[Walker et~al.(2003)]{Walker2003}
Walker, W.~E., Harremo\"es, P., Rotmans, J., van~der Sluijs, J.~P.,
van~Asselt, M.~B.~A., Janssen, P., \& Krayer~von~Krauss, M.~P. (2003).
Defining uncertainty: A conceptual basis for uncertainty management in
model-based decision support. \emph{Integrated Assessment}, 4(1), 5--17.

\bibitem[Wolpert(1996)]{Wolpert1996}
Wolpert, D.~H. (1996). The lack of a priori distinctions between
learning algorithms. \emph{Neural Computation}, 8(7), 1341--1390.

\bibitem[Wolpert \& Macready(1997)]{WolpertMacready1997}
Wolpert, D.~H., \& Macready, W.~G. (1997). No free lunch theorems for
optimization. \emph{IEEE Transactions on Evolutionary Computation},
1(1), 67--82.

\bibitem[Wynne(1992)]{Wynne1992}
Wynne, B. (1992). Uncertainty and environmental learning: Reconceiving
science and policy in the preventive paradigm. \emph{Global
Environmental Change}, 2(2), 111--127.

\end{thebibliography}
\end{document}